\documentclass[12pt]{article}
\usepackage{amsmath,amsfonts,amssymb}
\usepackage[pdftex,colorlinks=false]{hyperref}

\makeatletter
\def\numberbysection{\@addtoreset{equation}{section}
    \def\theequation{\thesection.\arabic{equation}}}
\makeatother

\numberbysection

\newcommand{\be}{\begin{eqnarray}}
\newcommand{\ee}{\end{eqnarray}}
\newcommand{\non}{\nonumber}

\newcommand{\tr}{\mathop{\rm tr}\nolimits}

\newcommand{\A}{\mathop{\cal{A}}\nolimits}
\newcommand{\B}{\mathop{\cal{B}}\nolimits}

\begin{document}

\begin{titlepage}
\strut\hfill UMTG--262
\vspace{.5in}
\begin{center}

\LARGE Two-loop test of the ${\cal N}=6$ Chern-Simons theory $S$-matrix\\
\vspace{1in}
\large Changrim Ahn \footnote{
       Department of Physics, Ewha Womans University,
       Seoul 120-750, South Korea} and
       Rafael I. Nepomechie \footnote{
       Physics Department, P.O. Box 248046, University of Miami,
       Coral Gables, FL 33124 USA}\\

\end{center}

\vspace{.5in}

\begin{abstract}
    Starting from the integrable two-loop spin-chain Hamiltonian
    describing the anomalous dimensions of scalar operators in the
    planar ${\cal N}=6$ superconformal Chern-Simons theory of ABJM, we
    perform a direct coordinate Bethe ansatz computation of the
    corresponding two-loop $S$-matrix.  The result matches with the
    weak-coupling limit of the scalar sector of the all-loop
    $S$-matrix which we have recently proposed.  In particular, we
    confirm that the scattering of $\A$ and $\B$ particles is
    reflectionless.  As a warm up, we first review the analogous
    computation of the one-loop $S$-matrix from the one-loop
    dilatation operator for the scalar sector of planar ${\cal N}=4$
    superconformal Yang-Mills theory, and compare the result with the
    all-loop $SU(2|2)^{2}$ $S$-matrix.
\end{abstract}

\end{titlepage}

\setcounter{footnote}{0}

\section{Introduction}\label{sec:intro}

Exact factorized $S$-matrices \cite{ZZ} play a key role in the
understanding of integrable models.  Planar four-dimensional ${\cal
N}=4$ superconformal Yang-Mills (YM) theory (and therefore, according
to the $AdS_{5}/CFT_{4}$ correspondence \cite{AdSCFT}, a certain type
IIB superstring theory on $AdS_{5}\times S^{5}$) is believed to be
integrable (see \cite{MZ1}-\cite{reviews} and references therein).  A
corresponding exact factorized $S$-matrix with $SU(2|2)^{2}$ symmetry
has been proposed (see \cite{St}-\cite{DHM} and references therein),
which leads \cite{Be, MM, dL} to the all-loop Bethe ansatz equations
(BAEs) \cite{BS2}.

Aharony, Bergman, Jafferis and Maldacena (ABJM) \cite{ABJM} recently
proposed an analogous $AdS_{4}/CFT_{3}$ correspondence relating planar
three-dimensional ${\cal N}=6$ superconformal Chern-Simons (CS) theory
to type IIA superstring theory on $AdS_4\times CP^3$.  Minahan and
Zarembo \cite{MZ2} subsequently found that the scalar sector of ${\cal
N}=6$ CS is integrable at the leading two-loop order, and proposed
two-loop BAEs for the full theory (see also \cite{BR}).  Moreover,
evidence for classical integrability of the dual string sigma model
(large-coupling limit) was discovered in \cite{AF2, Stef, GV1}.  On
the basis of these results, and assuming integrability to all orders,
Gromov and Vieira then conjectured all-loop BAEs \cite{GV2}.

Based on the symmetries and the spectrum of elementary excitations
\cite{MZ2, NT, GGY}, we proposed an exact factorized $AdS_{4}/CFT_{3}$
$S$-matrix \cite{AN1}.  As a check, we verified that this $S$-matrix
leads to the all-loop BAEs in \cite{GV2}.  An unusual feature of this
$S$-matrix is that the scattering of $\A$ and $\B$ particles is
reflectionless.  (A similar $S$-matrix which is not reflectionless
is not consistent with the known two-loop BAEs \cite{AN2}.)
For further related developments of the $AdS_{4}/CFT_{3}$
correspondence, see \cite{more} and references therein.

Considerable guesswork has entered into the above-mentioned all-loop
results.  While there is substantial evidence for the all-loop BAEs
and $S$-matrix in the well-studied $AdS_{5}/CFT_{4}$ case, the same
cannot be said for the rapidly-evolving $AdS_{4}/CFT_{3}$ case.  

In an effort to further check our proposed $S$-matrix, we perform here
a direct coordinate Bethe ansatz computation of the two-loop
$S$-matrix, starting from the integrable two-loop spin-chain
Hamiltonian describing the anomalous dimensions of scalar operators in
planar ${\cal N}=6$ CS \cite{MZ2}.  The result matches with the
weak-coupling limit of the scalar sector of our all-loop $S$-matrix
\cite{AN1}.  In particular, we confirm that the scattering of $\A$ and
$\B$ particles is reflectionless.  As a warm up, we first review the
analogous computation by Berenstein and V\'azquez \cite{BV} of the
one-loop $S$-matrix from the one-loop dilatation operator for the
scalar sector of planar ${\cal N}=4$ YM \cite{MZ1}, and compare the
result with the all-loop $SU(2|2)^{2}$ $S$-matrix.

The outline of this paper is as follows.  In Sec. \ref{sec:YM} we 
review the simpler case of ${\cal N}=4$ YM. In Sec. \ref{sec:CS} we 
analyze the ${\cal N}=6$ CS case, relegating most of the details of 
$\A-\B$ scattering to an appendix. We briefly discuss our results in 
Sec. \ref{sec:discussion}.

\section{One-loop $S$-matrix in the scalar sector of ${\cal N}=4$ YM}\label{sec:YM}

As is well known, ${\cal N}=4$ YM has six scalar fields $\Phi_{i}(x)$
($i=1, \ldots, 6$) in the adjoint representation of $SU(N)$.  It is
convenient to associate single-trace gauge-invariant scalar operators
with states of an $SO(6)$ quantum spin chain with $L$ sites,
\be
\tr \Phi_{i_{1}}(x) \cdots  \Phi_{i_{L}}(x) \quad \Leftrightarrow 
\quad | \Phi_{i_{1}}  \cdots  \Phi_{i_{L}} \rangle \,,
\ee 
where $\Phi_{i}$ on the RHS are 6-dimensional elementary vectors with
components $(\Phi_{i})_{j} = \delta_{i, j}$.  The one-loop anomalous
dimensions of these operators are described by the integrable $SO(6)$
quantum spin-chain Hamiltonian \cite{MZ1}
\be
\Gamma = \frac{\lambda}{8\pi^{2}} H\,, \qquad
H = \sum_{l=1}^{L}\left(1 - {\cal P}_{l, l+1} + \frac{1}{2}K_{l,l+1} 
\right) \,,
\ee
where $\lambda = g^{2}_{YM} N$ is the 't Hooft coupling, ${\cal P}$ is the
permutation operator, 
\be
{\cal P}\ \Phi_{i}\otimes \Phi_{j}  =  \Phi_{j}\otimes \Phi_{i}\,,
\ee
and the projector $K$ acts as
\be
K\ \Phi_{i}\otimes \Phi_{j}  = \delta_{i j} \left(\sum_{k=1}^{6} 
\Phi_{k}\otimes \Phi_{k} \right) \,.
\ee
It is convenient to define the complex combinations
\be
X=  \Phi_{1} + i  \Phi_{2} \,, \quad Y=\Phi_{3} + i  \Phi_{4} \,, 
\quad Z=\Phi_{5} + i  \Phi_{6} \,,
\ee
and to denote the corresponding complex conjugates with a bar, $\bar 
X =  \Phi_{1} - i  \Phi_{2}$, etc. For $\phi_{1}\,, \phi_{2} \in \{ 
X\,, \bar X\,, Y\,, \bar Y\,, Z\,, \bar Z\}$, 
\be
{\cal P}\ \phi_{1}\otimes \phi_{2}  =  \phi_{2}\otimes \phi_{1}\,,
\ee
and
\be
K\ \phi_{1} \otimes \phi_{2} = 
\left\{ \begin{array}{cc}
0 & \mbox{if } \phi_{1} \ne \bar \phi_{2} \\
X\otimes \bar X + \bar X \otimes X 
+ Y\otimes \bar Y + \bar Y \otimes Y 
+ Z\otimes \bar Z + \bar Z \otimes Z  & \mbox{if } \phi_{1} = \bar 
\phi_{2}
\end{array} \right. \,.
\ee

\subsection{Coordinate Bethe ansatz}

We take $| Z^{L} \rangle$ as the vacuum state, which evidently is an 
eigenstate of $H$ with zero energy.
One-particle excited states (``magnons'') with momentum $p$ are given by
\be
|\psi(p) \rangle_{\phi} = \sum_{x=1}^{L} e^{i p x} |x \rangle_{\phi} \,, 
\label{oneparticle}
\ee
where 
\be
|x \rangle_{\phi} = |\stackrel{\stackrel{1}{\downarrow}}{Z} \cdots Z 
\stackrel{\stackrel{x}{\downarrow}}{\phi} Z \cdots 
\stackrel{\stackrel{L}{\downarrow}}{Z} \rangle
\ee
is the
state obtained from the vacuum by replacing a single $Z$ at site $x$
with an ``impurity'' $\phi$, which can be either $X\,, \bar X\,, Y,
\bar Y$ (but not $\bar Z$, which can be regarded as a two-particle
bound state).  Indeed, one can easily check that (\ref{oneparticle})
is an eigenstate of $H$ with eigenvalue $E = \epsilon(p)$, where
\be
\epsilon(p) = 4 \sin^{2} (p/2) \,.
\label{epsilon}
\ee

In order to compute the two-particle $S$-matrix, we must construct 
all possible two-particle eigenstates. Let
\be
| x_{1}, x_{2} \rangle_{\phi_{1} \phi_{2}} = 
|\stackrel{\stackrel{1}{\downarrow}}{Z} \cdots  
\stackrel{\stackrel{x_{1}}{\downarrow}}{\phi_{1}}  
\cdots 
\stackrel{\stackrel{x_{2}}{\downarrow}}{\phi_{2}}  
\cdots 
\stackrel{\stackrel{L}{\downarrow}}{Z}\rangle
\ee
denote the state obtained from the vacuum by replacing the $Z$'s at
sites $x_{1}$ and $x_{2}$ with impurities $\phi_{1}$ and $\phi_{2}$,
respectively, where $x_{1} < x_{2}$.  Following Berenstein and
V\'azquez \cite{BV}, we distinguish the following three cases:

\begin{description}

\item[$\phi_{1} = \phi_{2}$:]\hfil

The case of two particles of the same type (i.e., $\phi_{1} = 
\phi_{2} \equiv \phi \in \{ X\,, \bar X\,, Y, \bar
Y\}$) is equivalent to the well-known case originally 
considered by Bethe in his seminal investigation of the Heisenberg 
model. (See, e.g., the review by Plefka in \cite{reviews}.) The 
two-particle eigenstates are given by
\be
|\psi \rangle = \sum_{x_{1}<x_{2}} 
f (x_{1}, x_{2})\, | x_{1}, x_{2} \rangle_{\phi  \phi} 
\ee
where
\be
f(x_{1}, x_{2}) = e^{i(p_{1} x_{1} + p_{2} x_{2})} + 
S(p_{2}\,, p_{1})\, e^{i(p_{2} x_{1} + p_{1} x_{2})} \,.
\label{fphiphi}
\ee
Indeed, these states satisfy
\be
H |\psi \rangle = E |\psi \rangle 
\label{EigenvalueProblem}
\ee
with
\be
E = \epsilon(p_{1}) + \epsilon(p_{2})  \,,
\label{2particleenergy}
\ee
where $\epsilon(p)$ is given by (\ref{epsilon}).  It also follows from
(\ref{EigenvalueProblem}) that the $S$-matrix for $\phi-\phi$
scattering is given by
\be
S(p_{2}\,, p_{1}) = \frac{u_{2} - u_{1} + 
i}{u_{2} - u_{1} - i}\,,
\label{Sphiphi}
\ee
where $u_{j} = u(p_{j})$ and 
\be
u(p) = \frac{1}{2} \cot (p/2) \,.
\ee

\item[$\phi_{1} \ne \bar{\phi}_{2}$:]\hfil

If the two particles are not of the same type, but $\phi_{1} \ne 
\bar{\phi}_{2}$, then the two-particle eigenstates are of the form
\be
|\psi \rangle = \sum_{x_{1}<x_{2}} \left\{
f_{\phi_{1} \phi_{2}}(x_{1}, x_{2})\, | x_{1}, x_{2} \rangle_{\phi_{1} 
\phi_{2}} + 
f_{\phi_{2} \phi_{1}}(x_{1}, x_{2})\, | x_{1}, x_{2} \rangle_{\phi_{2} 
\phi_{1}} \right\} \,,
\ee
where 
\be
f_{\phi_{i} \phi_{j}}(x_{1}, x_{2}) = 
A_{\phi_{i} \phi_{j}}(12)\, e^{i(p_{1} x_{1} + p_{2} x_{2})} 
+ A_{\phi_{i} \phi_{j}}(21)\, e^{i(p_{2} x_{1} + p_{1} x_{2})} \,.
\label{ansatz}
\ee
One finds \cite{BV}
\be
\left(
\begin{array}{c}
    A_{\phi_{1} \phi_{2}}(21) \\
    A_{\phi_{2} \phi_{1}}(21) 
\end{array} \right) = \left( \begin{array}{cc}
    R(p_{2}\,, p_{1}) & T(p_{2}\,, p_{1}) \\
    T(p_{2}\,, p_{1}) & R(p_{2}\,, p_{1}) 
    \end{array} \right) 
\left( 
\begin{array}{c}
    A_{\phi_{1} \phi_{2}}(12) \\
    A_{\phi_{2} \phi_{1}}(12) 
    \end{array} \right) \,, 
\ee     
where the transmission and reflection amplitudes are given by
\be
T(p_{2}\,, p_{1}) = \frac{u_{2}-u_{1}}{u_{2}-u_{1}-i} \,, \qquad
R(p_{2}\,, p_{1}) = \frac{i}{u_{2}-u_{1}-i} \,,
\label{Sphi1phi2}
\ee
respectively.

\item[$\phi_{1} = \bar{\phi}_{2}$:]\hfil

In the case $\phi_{1} = \bar{\phi}_{2}  \in \{ X\,, \bar X\,, Y, \bar
Y\}$, the two-particle eigenstates 
are given by
\be
|\psi \rangle &=& 
\sum_{x_{1}<x_{2}} \sum_{\phi = X, Y}\left\{
f_{\phi \bar \phi}(x_{1}, x_{2})\, | x_{1}, x_{2} \rangle_{\phi
\bar \phi} + 
f_{\bar \phi \phi}(x_{1}, x_{2})\, | x_{1}, x_{2} \rangle_{\bar \phi 
\phi} \right\} \non \\
& & + \sum_{x_{1}} f_{\bar Z}(x_{1}) | x_{1} \rangle_{\bar Z} \,,
\ee
where $f_{\phi_{i} \phi_{j}}(x_{1}, x_{2})$ are again given by 
(\ref{ansatz}), and 
\be
f_{\bar Z}(x_{1}) = A_{\bar Z}\, e^{i(p_{1}+p_{2}) x_{1}}\,.
\ee
One finds \cite{BV}
\be
\hspace{-0.4in} \left(
\begin{array}{c}
    A_{X \bar X}(21) \\
    A_{\bar X X}(21) \\
    A_{Y \bar Y}(21) \\
    A_{\bar Y Y}(21)    
\end{array} \right) = \left( \begin{array}{cccc}
    R(p_{2}\,, p_{1}) & T(p_{2}\,, p_{1}) & S(p_{2}\,, p_{1}) & S(p_{2}\,, p_{1})\\
    T(p_{2}\,, p_{1}) & R(p_{2}\,, p_{1}) & S(p_{2}\,, p_{1}) & S(p_{2}\,, p_{1})\\
    S(p_{2}\,, p_{1}) & S(p_{2}\,, p_{1}) & R(p_{2}\,, p_{1}) & T(p_{2}\,, p_{1})\\
    S(p_{2}\,, p_{1}) & S(p_{2}\,, p_{1}) & T(p_{2}\,, p_{1}) & R(p_{2}\,, p_{1})        
    \end{array} \right) 
\left( 
\begin{array}{c}
    A_{X \bar X}(12) \\
    A_{\bar X X}(12) \\
    A_{Y \bar Y}(12) \\
    A_{\bar Y Y}(12)     
    \end{array} \right)  \,,
\ee     
where 
\be
T(p_{2}\,, p_{1}) &=& 
\frac{(u_{2}-u_{1})^{2}}{(u_{2}-u_{1}-i)(u_{2}-u_{1}+i)} \,, \non \\
R(p_{2}\,, p_{1}) &=& 
\frac{-1}{(u_{2}-u_{1}-i)(u_{2}-u_{1}+i)} \,, \non \\
S(p_{2}\,, p_{1}) &=& 
\frac{-i(u_{2}-u_{1})}{(u_{2}-u_{1}-i)(u_{2}-u_{1}+i)} \,.
\label{Sphibarphi}
\ee

\end{description}

\subsection{Comparison with the all-loop $S$-matrix}

We now wish to compare the above scattering amplitudes with the
weak-coupling limit of the all-loop $SU(2|2)\otimes SU(2|2)$
$S$-matrix \cite{Be}-\cite{DHM}.  This check has not (to our
knowledge) been presented elsewhere, and will serve as a useful guide
for the ${\cal N}=6$ CS case.  It is convenient to express the latter
in terms of two mutually commuting sets of Zamolodchikov-Faddeev
operators $A_{i}^{\dagger}(p)\,, \tilde A_{i}^{\dagger}(p)$ ($i=1,
\ldots, 4$),
\be
A_{i}^{\dagger}(p_{1})\, A_{j}^{\dagger}(p_{2}) &=&
\sum_{i', j'} S_{0}(p_{1}, p_{2})\,  \widehat S_{i\, j}^{i' j'}(p_{1}, p_{2})\, 
A_{j'}^{\dagger}(p_{2})\, A_{i'}^{\dagger}(p_{1}) \,, \non \\ 
\tilde A_{i}^{\dagger}(p_{1})\, \tilde A_{j}^{\dagger}(p_{2}) &=&
\sum_{i', j'} S_{0}(p_{1}, p_{2})\,  \widehat S_{i\, j}^{i' j'}(p_{1}, p_{2})\, 
\tilde A_{j'}^{\dagger}(p_{2})\, \tilde A_{i'}^{\dagger}(p_{1}) \,, 
\non \\
A_{i}^{\dagger}(p_{1})\, \tilde A_{j}^{\dagger}(p_{2}) &=& 
\tilde A_{j}^{\dagger}(p_{2})\, A_{i}^{\dagger}(p_{1}) \,.
\label{bulkS}
\ee
We identify the scalar one-particle states as follows,
\be
X(p) &=& A_{1}^\dagger(p)\, \tilde A_{2}^\dagger(p) \,, \quad
\bar X(p) = A_{2}^\dagger(p)\, \tilde A_{1}^\dagger(p) \,, \non \\
Y(p) &=& A_{2}^\dagger(p)\, \tilde A_{2}^\dagger(p) \,, \quad
\bar Y(p) = A_{1}^\dagger(p)\, \tilde A_{1}^\dagger(p) \,.
\label{identification}
\ee
The only non-vanishing amplitudes in the scalar sector are
\be
\widehat S_{a\, a}^{a\, a}(p_{1}, p_{2}) =  A\,, \quad 
\widehat S_{a\, b}^{a\, b}(p_{1}, p_{2}) = \frac{1}{2}(A-B)\,, \quad 
\widehat S_{a\, b}^{b\, a}(p_{1}, p_{2}) = \frac{1}{2}(A+B) \,,
\label{Sscalarsector}
\ee
where $a\,, b \in \{1\,, 2\}$ with $a \ne b$. Here
\be
A &=& \frac{x^{-}_{2}-x^{+}_{1}}{x^{+}_{2}-x^{-}_{1}} \,, \non \\
B &=&-\left[\frac{x^{-}_{2}-x^{+}_{1}}{x^{+}_{2}-x^{-}_{1}}+
2\frac{(x^{-}_{1}-x^{+}_{1})(x^{-}_{2}-x^{+}_{2})(x^{-}_{2}+x^{+}_{1})}
{(x^{-}_{1}-x^{+}_{2})(x^{-}_{1}x^{-}_{2}-x^{+}_{1}x^{+}_{2})}\right]\,,
\label{ABamplitudes}
\ee
where $x_{i}^{\pm} = x(p_{i})^{\pm}$ with
\be
\frac{x^{+}}{x^{-}} = e^{i p}\,, \quad 
x^{+}+\frac{1}{x^{+}}-x^{-}-\frac{1}{x^{-}} = \frac{i}{g} \,,
\label{xpm}
\ee 
and $g=\sqrt{\lambda}/(4\pi)$.
Moreover, the scalar factor is given by
\be
S_{0}(p_{1}\,, p_{2})^{2} = \frac{x^{-}_{1}-x^{+}_{2}}{x^{+}_{1}-x^{-}_{2}}
\frac{1-\frac{1}{x^{+}_{1}x^{-}_{2}}}{1-\frac{1}{x^{-}_{1}x^{+}_{2}}}
\sigma(p_{1}\,, p_{2})^{2} \,,
\label{AFrelation}
\ee
where $\sigma(p_{1}\,, p_{2})$ is the BES dressing factor \cite{BES, 
DHM}. In the weak-coupling ($g \rightarrow 0$) limit, 
\be
x^{\pm} \rightarrow \frac{1}{g}\left(u \pm \frac{i}{2}\right) \,.
\label{weakcoupling1}
\ee
Therefore
\be
A \rightarrow \frac{u_{1} - u_{2} + i}{u_{1} - u_{2} - i} \,, \qquad
B \rightarrow -1 \,,
\label{weakcoupling2}
\ee
and 
\be
S_{0}^{2} \rightarrow \frac{u_{1} - u_{2} - i}{u_{1} - u_{2} + i} \,,
\label{weakcoupling3}
\ee 
since $\sigma(p_{1}\,, p_{2}) \rightarrow 1$.

For two particles of the same type, the scattering amplitude is evidently given 
by
\be
S(p_{1}, p_{2}) \equiv 
\left( S_{0}(p_{1}, p_{2})\, \widehat S_{a\, a}^{a\, a}(p_{1}, p_{2}) \right)^{2} 
= S_{0}^{2}\, A^{2} \rightarrow \frac{u_{1} - u_{2} + i}{u_{1} - u_{2} - i} \,,
\label{N4YMsame}
\ee
in agreement with (\ref{Sphiphi}). 

We now consider the case $\phi_{1} \ne \bar \phi_{2}$, e.g.,
\be
X(p_{1})\, Y(p_{2}) = T(p_{1}, p_{2})\,  Y(p_{2})\, X(p_{1}) + 
R(p_{1}, p_{2})\,  X(p_{2})\, Y(p_{1}) \,.
\ee
It follows from (\ref{bulkS})-(\ref{Sscalarsector}) and
(\ref{weakcoupling2}), (\ref{weakcoupling3}) that
\be
T(p_{1}, p_{2}) &=& \frac{1}{2}S_{0}^{2} A(A-B) \rightarrow 
\frac{u_{1}-u_{2}}{u_{1}-u_{2}-i}  \,, \non \\
R(p_{1}, p_{2}) &=& \frac{1}{2}S_{0}^{2} A(A+B) \rightarrow 
\frac{i}{u_{1}-u_{2}-i}\,, 
\label{N4YMsame2}
\ee
in agreement with (\ref{Sphi1phi2}).

Finally, we consider the case $\phi_{1} = \bar \phi_{2}$, e.g.,
\be
X(p_{1})\, \bar X(p_{2}) &=& T(p_{1}, p_{2})\,  \bar X(p_{2})\, X(p_{1}) + 
R(p_{1}, p_{2})\,  X(p_{2})\, \bar X(p_{1}) \non \\
&+& S(p_{1}, p_{2})\,  Y(p_{2})\, \bar Y(p_{1}) + 
S(p_{1}, p_{2})\, \bar Y(p_{2})\, Y(p_{1})
\label{ZFXbX}
\,.
\ee
It follows from (\ref{bulkS})-(\ref{Sscalarsector}) and
(\ref{weakcoupling2}), (\ref{weakcoupling3}) that
\be
T(p_{1}, p_{2}) &=& \frac{1}{4}S_{0}^{2} (A-B)^{2} \rightarrow 
\frac{(u_{1}-u_{2})^{2}}{(u_{1}-u_{2}-i)(u_{1}-u_{2}+i)}  \,, \non \\
R(p_{1}, p_{2}) &=& \frac{1}{4}S_{0}^{2} (A+B)^{2} \rightarrow 
\frac{-1}{(u_{1}-u_{2}-i)(u_{1}-u_{2}+i)}\,, \non \\
S(p_{1}, p_{2}) &=& \frac{1}{4}S_{0}^{2} (A-B)(A+B) \rightarrow 
\frac{i(u_{1}-u_{2})}{(u_{1}-u_{2}-i)(u_{1}-u_{2}+i)}\,,
\ee
in agreement with (\ref{Sphibarphi}).\footnote{There is a sign 
discrepancy in $S(p_{1}, p_{2})$ which perhaps can be reconciled by a 
gauge transformation in (\ref{ZFXbX}), e.g., $Y \rightarrow -Y$ while 
leaving others unchanged.}

In short, the all-loop $AdS_{5}/CFT_{4}$ $S$-matrix correctly
reproduces the ${\cal N}=4$ YM one-loop scalar-sector scattering
amplitudes, as expected.  In the next section, we perform a similar
check of the $AdS_{4}/CFT_{3}$ $S$-matrix.

\section{Two-loop $S$-matrix in the scalar sector of ${\cal N}=6$ CS}\label{sec:CS}

The ${\cal N}=6$ CS theory \cite{ABJM} has a pair of scalar fields
$A_{i}(x)$ ($i = 1, 2$) in the bifundamental representation $({\bf N},
{\bf \bar N})$ of the $SU(N) \times SU(N)$ gauge group, and another
pair of scalar fields $B_{i}(x)$ ($i = 1, 2$) in the conjugate
representation $({\bf \bar N}, {\bf N})$.  These fields can be grouped
into $SU(4)$ multiplets $Y^{A}(x)$,
\be
Y^{A} = (A_{1}\,, A_{2}\,, B_{1}^{\dagger}\,, B_{2}^{\dagger})\,, 
\qquad
Y_{A}^{\dagger} = (A_{1}^{\dagger}\,, A_{2}^{\dagger}\,, B_{1}\,, B_{2})\,.
\ee
Following \cite{MZ2}, we associate single-trace gauge-invariant scalar
operators with states of an alternating $SU(4)$ quantum spin chain
with $2L$ sites,
\be
\tr Y^{A_{1}}(x)\, Y_{B_{1}}^{\dagger}(x) \cdots  Y^{A_{L}}(x)\, 
Y_{B_{L}}^{\dagger}(x) \quad \Leftrightarrow 
\quad | Y^{A_{1}}\, Y_{B_{1}}^{\dagger} \cdots  Y^{A_{L}}\, 
Y_{B_{L}}^{\dagger} \rangle \,,
\ee 
where $Y^{A}$ on the RHS are 4-dimensional elementary vectors with
components $(Y^{A})_{j} = \delta_{A, j}$.  The two-loop anomalous
dimensions of these operators are described by the integrable 
alternating $SU(4)$ quantum spin-chain Hamiltonian \cite{MZ2}
\be
\Gamma = \lambda^{2} H\,, \qquad
H = \sum_{l=1}^{2L}\left(1 - {\cal P}_{l, l+2} + \frac{1}{2}\{ 
K_{l,l+1} \,, {\cal P}_{l, l+2} \}
\right) \,,
\label{HCS}
\ee
where $\lambda = N/k$ is the 't Hooft coupling\footnote{The action
has two $SU(N)$ Chern-Simons terms with integer levels $k$ and $-k$, 
respectively.}, ${\cal P}$ is the permutation operator, 
and the projector $K$ acts as
\be
K\ Y^{A}\otimes Y_{B}^{\dagger}  =  \delta^{A}_{B} \sum_{C=1}^{4} 
Y^{C}\otimes Y_{C}^{\dagger} 
\,, \qquad 
K\ Y_{B}^{\dagger} \otimes Y^{A}  =  \delta^{A}_{B} \sum_{C=1}^{4}
Y_{C}^{\dagger}\otimes  Y^{C}\,.
\ee
That is,
\be
K\, A_{i} \otimes A_{j}^{\dagger} &=& K\, B_{i}^{\dagger} \otimes B_{j} 
= \delta_{i j}\sum_{k=1}^{2}  \left( A_{k} \otimes A_{k}^{\dagger} + 
B_{k}^{\dagger} \otimes B_{k} \right) \,, \non \\
K\, A_{i}^{\dagger} \otimes A_{j} &=& K\, B_{i} \otimes B_{j}^{\dagger} 
= \delta_{i j}\sum_{k=1}^{2}  \left( A_{k}^{\dagger} \otimes A_{k} + 
B_{k} \otimes B_{k}^{\dagger} \right) \,, \non \\
K\, A_{i} \otimes B_{j} &=& K\, B_{i} \otimes A_{j} = K\, 
A_{i}^{\dagger} \otimes B_{j}^{\dagger} = K\, B_{i}^{\dagger} \otimes 
A_{j}^{\dagger} = 0 \,.
\ee

\subsection{Coordinate Bethe ansatz}

Following \cite{NT, GGY}, we take the state with $L$ pairs of $(A_{1} B_{1})$, i.e.,
\be
| (A_{1} B_{1})^{L} \rangle
\ee 
as the vacuum state, which evidently is an eigenstate of $H$ with zero
energy. It is convenient to label the $(A_{1} B_{1})$ pairs by $x \in \{1, \ldots, L \}$.
There are two types
of one-particle excited states with momentum $p$, called ``$\A$-particles'' and 
``$\B$-particles.'' The former are given by
\be
|\psi(p) \rangle_{\phi}^{\A} = \sum_{x=1}^{L} e^{i p x} |x 
\rangle_{\phi}^{\A} \,, 
\label{Aparticle}
\ee
where 
\be
|x \rangle_{\phi}^{\A} = |\stackrel{\stackrel{1}{\downarrow}}{(A_{1}B_{1})} 
\cdots 
\stackrel{\stackrel{x}{\downarrow}}{(\phi B_{1})}
\cdots 
\stackrel{\stackrel{L}{\downarrow}}{(A_{1}B_{1})}  \rangle
\ee
is the state obtained from the vacuum by replacing the $A_{1}$ from
pair $x$ with an ``impurity'' $\phi$, which
can be either $A_{2}$ or $B_{2}^{\dagger}$ (but not $B_{1}^{\dagger}$,
which can be regarded as a two-particle bound state).  Similarly, the
``$\B$-particles'' are given by
\be
|\psi(p) \rangle_{\phi}^{\B} = \sum_{x=1}^{L} e^{i p x} |x 
\rangle_{\phi}^{\B} \,, 
\label{Bparticle}
\ee
where 
\be
|x \rangle_{\phi}^{\B} = |\stackrel{\stackrel{1}{\downarrow}}{(A_{1}B_{1})} 
\cdots 
\stackrel{\stackrel{x}{\downarrow}}{(A_{1} \phi)}
\cdots 
\stackrel{\stackrel{L}{\downarrow}}{(A_{1}B_{1})} \rangle
\ee
is the state obtained from the vacuum by replacing the $B_{1}$ from
pair $x$ with an ``impurity'' $\phi$, which can
be either $A_{2}^{\dagger}$ or $B_{2}$ (but not $A_{1}^{\dagger}$, which
can be regarded as a two-particle bound state). Indeed, both 
(\ref{Aparticle}) and (\ref{Bparticle}) are eigenstates of
$H$ with eigenvalue $E = \epsilon(p)$, where $\epsilon(p)$ is given 
by (\ref{epsilon}).

In order to compute the two-particle $S$-matrix, we must construct 
all possible two-particle eigenstates. 

\subsubsection{$\A-\A$ scattering}

Let
\be
| x_{1}, x_{2} \rangle_{\phi_{1} \phi_{2}}^{\A\A} = 
|\stackrel{\stackrel{1}{\downarrow}}{(A_{1}B_{1})}  
\cdots  
\stackrel{\stackrel{x_{1}}{\downarrow}}{(\phi_{1} B_{1})}
\cdots 
\stackrel{\stackrel{x_{2}}{\downarrow}}{(\phi_{2} B_{1})}  
\cdots 
\stackrel{\stackrel{L}{\downarrow}}{(A_{1}B_{1})}  \rangle
\ee
denote the state obtained from the vacuum by replacing the $A_{1}$'s 
from pairs $x_{1}$ and $x_{2}$ with impurities $\phi_{1}$ and $\phi_{2}$,
respectively, where $x_{1} < x_{2}$ and $\phi_{i} \in \{ A_{2}\,, B_{2}^{\dagger} \}$. 
We distinguish two cases:

\begin{description}

\item[$\phi_{1} = \phi_{2}$:]\hfil

The case of two $\A$-particles of the same type (i.e., $\phi_{1} = 
\phi_{2} \equiv \phi \in \{ A_{2}\,, B_{2}^{\dagger} \}$) is again 
the same as in the Heisenberg model.  The 
two-particle eigenstates are given by
\be
|\psi \rangle = \sum_{x_{1}<x_{2}} 
f(x_{1}, x_{2})\, | x_{1}, x_{2} \rangle_{\phi  
\phi}^{\A\A} 
\label{AAeigenstates1}
\ee
where $f(x_{1}, x_{2})$ is given by (\ref{fphiphi}).
These states have energy (\ref{2particleenergy}), 
and the $S$-matrix is again given by (\ref{Sphiphi}), 
\be
S(p_{2}\,, p_{1}) = \frac{u_{2} - u_{1} + 
i}{u_{2} - u_{1} - i}\,.
\label{SAAphiphi}
\ee

\item[$\phi_{1} \ne \phi_{2}$:]\hfil

If the two $\A$-particles are not of the same type (e.g., $\phi_{1} = 
A_{2}\,, \phi_{2} =B_{2}^{\dagger}$), then the
two-particle eigenstates are of the form
\be
|\psi \rangle = \sum_{x_{1}<x_{2}} \left\{
f_{\phi_{1} \phi_{2}}(x_{1}, x_{2})\, | x_{1}, x_{2} \rangle_{\phi_{1} 
\phi_{2}}^{\A\A} + 
f_{\phi_{2} \phi_{1}}(x_{1}, x_{2})\, | x_{1}, x_{2} \rangle_{\phi_{2} 
\phi_{1}}^{\A\A} \right\} \,,
\label{AAeigenstates2}
\ee
where $f_{\phi_{i} \phi_{j}}(x_{1}, x_{2})$ is again given by
(\ref{ansatz}).  Since $K$ on these states is zero, the $S$-matrix is
again given by (\ref{Sphi1phi2}),
\be
T(p_{2}\,, p_{1}) = \frac{u_{2}-u_{1}}{u_{2}-u_{1}-i} \,, \qquad
R(p_{2}\,, p_{1}) = \frac{i}{u_{2}-u_{1}-i} \,.
\label{SAAphi1phi2}
\ee

\end{description}

\subsubsection{$\B-\B$ scattering}\label{subsubsec:coordBB}

Let
\be
| x_{1}, x_{2} \rangle_{\phi_{1} \phi_{2}}^{\B\B} = 
|\stackrel{\stackrel{1}{\downarrow}}{(A_{1}B_{1})}  
\cdots 
\stackrel{\stackrel{x_{1}}{\downarrow}}{(A_{1} \phi_{1})}
\cdots 
\stackrel{\stackrel{x_{2}}{\downarrow}}{(A_{1} \phi_{2})} 
\cdots 
\stackrel{\stackrel{L}{\downarrow}}{(A_{1}B_{1})}  \rangle
\ee
denote the state obtained from the vacuum by replacing the $B_{1}$'s 
from pairs $x_{1}$ and $x_{2}$ with impurities $\phi_{1}$ and $\phi_{2}$,
respectively, where $x_{1} < x_{2}$  and $\phi_{i} \in \{ A_{2}^{\dagger}\,, B_{2} \}$.
The eigenstates with two $\B$-particle are given by expressions 
similar to those with two $\A$-particles (namely, 
(\ref{AAeigenstates1}) and (\ref{AAeigenstates2}) with 
$| x_{1}, x_{2} \rangle_{\phi_{i} \phi_{j}}^{\A\A} \leftrightarrow 
| x_{1}, x_{2} \rangle_{\phi_{i} \phi_{j}}^{\B\B}$), and we obtain the 
same results (\ref{SAAphiphi}), (\ref{SAAphi1phi2}) for the scattering amplitudes .

\subsubsection{$\A-\B$ scattering}\label{subsubsec:coordAB}

In order to analyze $\A-\B$ scattering, we define the states
\be
| x_{1}, x_{2} \rangle_{\phi_{1} \phi_{2}}^{\A\B} &=& 
|\stackrel{\stackrel{1}{\downarrow}}{(A_{1}B_{1})}
\cdots  
\stackrel{\stackrel{x_{1}}{\downarrow}}{(\phi_{1} B_{1})}
\cdots 
\stackrel{\stackrel{x_{2}}{\downarrow}}{(A_{1} \phi_{2})}
\cdots 
\stackrel{\stackrel{L}{\downarrow}}{(A_{1}B_{1})}  \rangle \,, \non \\
| x_{1}, x_{2} \rangle_{\phi_{2} \phi_{1}}^{\A\B} &=& 
|\stackrel{\stackrel{1}{\downarrow}}{(A_{1}B_{1})} 
\cdots
\stackrel{\stackrel{x_{1}}{\downarrow}}{(A_{1} \phi_{2}) }
\cdots
\stackrel{\stackrel{x_{2}}{\downarrow}}{(\phi_{1} B_{1})} 
\cdots 
\stackrel{\stackrel{L}{\downarrow}}{(A_{1}B_{1})}  \rangle \,,
\label{ABstates}
\ee
where $x_{1} < x_{2}$ and $\phi_{1} \in \{ A_{2}\,, B_{2}^{\dagger} \}$, 
$\phi_{2} \in \{ A_{2}^{\dagger}\,, B_{2} \}$. We distinguish two 
cases:

\begin{description}

\item[$\phi_{1} \ne \phi_{2}^{\dagger}$:]\hfil

If $\phi_{1} \ne \phi_{2}^{\dagger}$ (e.g., $\phi_{1} = 
A_{2}\,, \phi_{2} =B_{2}$), then $K$ on the states (\ref{ABstates}) is zero.
As noted in \cite{MZ2}, we are left with two decoupled $SU(2)$ chains 
on the even and odd sites.  Hence, there is trivial scattering 
between $\A$ 
and $\B$ particles.

\item[$\phi_{1} = \phi_{2}^{\dagger}$:]\hfil

If $\phi_{1} = \phi_{2}^{\dagger}$ (e.g., $\phi_{1} = 
A_{2}\,, \phi_{2} =A_{2}^{\dagger}$),
then the eigenstates are given by
\be
|\psi \rangle &=& 
\sum_{x_{1}<x_{2}} \sum_{\phi = A_{2}, B_{2}^{\dagger}}\left\{
f_{\phi \phi^{\dagger}}(x_{1}, x_{2})\, | x_{1}, x_{2} \rangle_{\phi
\phi^{\dagger}}^{\A\B} + 
f_{\phi^{\dagger} \phi}(x_{1}, x_{2})\, | x_{1}, x_{2} \rangle_{\phi^{\dagger} 
\phi}^{\A\B} \right\} \non \\
& & + \sum_{x_{1}} \sum_{k=1}^{2} \left\{
f_{A_{k} A_{k}^{\dagger}}(x_{1}) | x_{1} \rangle_{A_{k} A_{k}^{\dagger}} 
+ f_{B_{k}^{\dagger} B_{k}}(x_{1}) | x_{1} \rangle_{B_{k}^{\dagger} 
B_{k}} \right\} \,,
\label{ABeigenket}
\ee
where 
\be
|x \rangle_{\phi_{i} \phi_{j}} = 
|\stackrel{\stackrel{1}{\downarrow}}{(A_{1}B_{1})}
\cdots 
\stackrel{\stackrel{x}{\downarrow}}{(\phi_{i} \phi_{j})}
\cdots 
\stackrel{\stackrel{L}{\downarrow}}{(A_{1}B_{1})} \rangle
\ee
is the state obtained from the vacuum by replacing the $(A_{1} B_{1})$
pair at $x$ with $(\phi_{i} \phi_{j})$.  We assume that $f_{\phi_{i}
\phi_{j}}(x_{1}, x_{2})$ are again given by (\ref{ansatz}), and
\be
f_{\phi_{i} \phi_{j}}(x_{1}) = A_{\phi_{i} \phi_{j}}\, e^{i(p_{1}+p_{2}) x_{1}}\,.
\ee
After a lengthy computation (see 
the Appendix for further details), we find 
\be
\hspace{-0.4in} \left(
\begin{array}{c}
    A_{A_{2} A_{2}^{\dagger}}(21) \\
    A_{A_{2}^{\dagger} A_{2}}(21) \\
    A_{B_{2}^{\dagger} B_{2}}(21) \\
    A_{B_{2} B_{2}^{\dagger}}(21)    
\end{array} \right) = \left( \begin{array}{cccc}
    0 & T(p_{2}\,, p_{1}) & 0 & S(p_{2}\,, p_{1})\\
    T(p_{2}\,, p_{1}) & 0 & S(p_{2}\,, p_{1}) & 0\\
    0 & S(p_{2}\,, p_{1}) & 0 & T(p_{2}\,, p_{1})\\
    S(p_{2}\,, p_{1}) & 0 & T(p_{2}\,, p_{1}) & 0        
    \end{array} \right) 
\left( 
\begin{array}{c}
    A_{A_{2} A_{2}^{\dagger}}(12) \\
    A_{A_{2}^{\dagger} A_{2}}(12) \\
    A_{B_{2}^{\dagger} B_{2}}(12) \\
    A_{B_{2} B_{2}^{\dagger}}(12)   
    \end{array} \right) 
\label{inout}
\ee     
where 
\be
T(p_{2}\,, p_{1}) = 
\frac{u_{1}-u_{2}}{u_{1}-u_{2}-i} \,, \qquad 
S(p_{2}\,, p_{1}) =
\frac{i}{u_{1}-u_{2}-i} \,.
\label{SABphibarphi}
\ee
Note that the scattering is reflectionless.

Similar results can be obtained for $\B-\A$ scattering.

\end{description}

\subsection{Comparison with the all-loop $S$-matrix}

We now wish to compare the above scattering amplitudes with the
weak-coupling limit of the all-loop $SU(2|2)$ $S$-matrix \cite{AN1}.
It is convenient to express the latter in terms of two sets of 
Zamolodchikov-Faddeev operators $\A_{i}^{\dagger}(p)$, 
$\B_{i}^{\dagger}(p)$ ($i=1, \ldots, 4$) corresponding to the $\A$,  
$\B$ particles, respectively,
\be
\A_{i}^{\dagger}(p_{1})\, \A_{j}^{\dagger}(p_{2}) &=&
\sum_{i', j'} S_{0}(p_{1}, p_{2})\,  \widehat S_{i\, j}^{i' j'}(p_{1}, p_{2})\, 
\A_{j'}^{\dagger}(p_{2})\, \A_{i'}^{\dagger}(p_{1}) \,, \label{SAA} \\ 
\B_{i}^{\dagger}(p_{1})\, \B_{j}^{\dagger}(p_{2}) &=&
\sum_{i', j'} S_{0}(p_{1}, p_{2})\,  \widehat S_{i\, j}^{i' j'}(p_{1}, p_{2})\, 
\B_{j'}^{\dagger}(p_{2})\, \B_{i'}^{\dagger}(p_{1}) \,, 
\label{SBB} \\
\A_{i}^{\dagger}(p_{1})\, \B_{j}^{\dagger}(p_{2}) &=&
\sum_{i', j'} \tilde S_{0}(p_{1}, p_{2})\,  \widehat S_{i\, j}^{i' j'}(p_{1}, p_{2})\, 
\B_{j'}^{\dagger}(p_{2})\, \A_{i'}^{\dagger}(p_{1}) \,, \label{SAB} \\ 
\B_{i}^{\dagger}(p_{1})\, \A_{j}^{\dagger}(p_{2}) &=&
\sum_{i', j'} \tilde S_{0}(p_{1}, p_{2})\,  \widehat S_{i\, j}^{i' j'}(p_{1}, p_{2})\, 
\A_{j'}^{\dagger}(p_{2})\, \B_{i'}^{\dagger}(p_{1}) \,.
\label{SBA}
\ee
The absence of 
$\A_{j'}^{\dagger}(p_{2})\, \B_{i'}^{\dagger}(p_{1})$ terms
on the RHS of (\ref{SAB}) (and similarly, of 
$\B_{j'}^{\dagger}(p_{2})\, \A_{i'}^{\dagger}(p_{1})$ terms on the 
RHS of  (\ref{SBA})) means that the scattering is reflectionless.

We identify the scalar one-particle states as follows,
\be
\A_{1}^{\dagger}(p) |0 \rangle &=& \sum_{x} e^{i p x} |x \rangle_{A_{2}}^{\A} \,, 
\qquad
\A_{2}^{\dagger}(p) |0 \rangle = \sum_{x} e^{i p x} |x 
\rangle_{B_{2}^{\dagger}}^{\A} \,, \non \\
\B_{1}^{\dagger}(p) |0 \rangle &=& \sum_{x} e^{i p x} |x 
\rangle_{B_{2}}^{\B} \,, 
\qquad
\B_{2}^{\dagger}(p) |0 \rangle = \sum_{x} e^{i p x} |x 
\rangle_{A_{2}^{\dagger}}^{\B} \,.
\ee 
The $SU(2|2)$ $S$-matrix elements $\widehat S_{i\, j}^{i' j'}(p_{1}, 
p_{2})$ are the same as before (\ref{Sscalarsector}), 
(\ref{ABamplitudes}), where $x^{\pm}$ satisfy (\ref{xpm}) and \cite{NT, GGY, GHO}
\be
g=h(\lambda) \,,
\label{gvalue}
\ee
with $h(\lambda) \sim \lambda$ for small $\lambda$, and
$h(\lambda) \sim \sqrt{\lambda/2}$ for large $\lambda$.
The scalar factors are given by (cf. (\ref{AFrelation}))
\be
S_{0}(p_{1}\,, p_{2}) =
\frac{1-\frac{1}{x^{+}_{1}x^{-}_{2}}}{1-\frac{1}{x^{-}_{1}x^{+}_{2}}}
\sigma(p_{1}\,, p_{2}) \,, \qquad
\tilde S_{0}(p_{1}\,, p_{2}) =
\frac{x^{-}_{1}-x^{+}_{2}}{x^{+}_{1}-x^{-}_{2}}
\sigma(p_{1}\,, p_{2}) \,.
\label{S0values}
\ee
In the weak-coupling ($g \rightarrow 0$) limit, 
\be
S_{0} \rightarrow 1 \,, \qquad 
\tilde S_{0} \rightarrow \frac{u_{1} - u_{2} - i}{u_{1} - u_{2} + i} 
\,.
\ee 

\subsubsection{$\A-\A$ scattering}

For two $\A$ particles of the same type (i.e., both $\A_{a}$ with 
$a \in \{1, 2\}$), the scattering amplitude is 
evidently given by
\be
S(p_{1}, p_{2}) \equiv  
S_{0}(p_{1}, p_{2})\, \widehat S_{a\, a}^{a\, a}(p_{1}, p_{2}) 
= S_{0}\, A \rightarrow \frac{u_{1} - u_{2} + i}{u_{1} - u_{2} - i} \,,
\ee
in agreement with (\ref{SAAphiphi}).  Although the same expression
also appears in the ${\cal N}=4$ YM case (\ref{N4YMsame}), note that
the latter follows from the all-loop $S$-matrix (\ref{bulkS}) in a
rather different way.

For two $\A$ particles of different type (i.e., $\A_{a}$ and $\A_{b}$
with $a, b \in \{1, 2\}$ and $a \ne b$), it follows from (\ref{SAA}),
(\ref{Sscalarsector}), (\ref{weakcoupling2}) that
\be
\A_{a}^{\dagger}(p_{1})\, \A_{b}^{\dagger}(p_{2}) &=&
T(p_{1}, p_{2})\,  \A_{b}^{\dagger}(p_{2})\, \A_{a}^{\dagger}(p_{1}) +
R(p_{1}, p_{2})\,  \A_{a}^{\dagger}(p_{2})\, \A_{b}^{\dagger}(p_{1}) 
\,,
\ee
where
\be
T(p_{1}, p_{2}) &=& \frac{1}{2}S_{0} (A-B) \rightarrow 
\frac{u_{1}-u_{2}}{u_{1}-u_{2}-i}  \,, \non \\
R(p_{1}, p_{2}) &=& \frac{1}{2}S_{0} (A+B) \rightarrow 
\frac{i}{u_{1}-u_{2}-i}\,, 
\ee
in agreement with (\ref{SAAphi1phi2}). Again, the same expressions 
arise in the ${\cal N}=4$ YM case (\ref{N4YMsame2}) in a different 
way.

\subsubsection{$\B-\B$ scattering}

According to (\ref{SBB}), the $\B-\B$ and $\A-\A$ scattering
amplitudes are equal, in agreement with the results from Sec.
\ref{subsubsec:coordBB}.

\subsubsection{$\A-\B$ scattering}

According to (\ref{SAB}), the $\A_{a}-\B_{a}$ scattering 
amplitude is 
\be
\tilde S_{0}(p_{1}, p_{2})\, \widehat S_{a\, a}^{a\, a}(p_{1}, p_{2}) 
= \tilde S_{0}\, A \rightarrow 1 \,,
\ee
in agreement with the results from Sec.\ref{subsubsec:coordAB} for the
case $\phi_{1} \ne \phi_{2}^{\dagger}$.  Note that the scalar factor $
\tilde S_{0}$ (\ref{S0values}) is essential for obtaining this result.

For $\A_{a}-\B_{b}$ scattering (with $a, b \in \{1, 2\}$ and $a \ne 
b$), it follows from (\ref{SAB}) that
\be
\A_{a}^{\dagger}(p_{1})\, \B_{b}^{\dagger}(p_{2}) &=&
T(p_{1}, p_{2})\,  \B_{b}^{\dagger}(p_{2})\, \A_{a}^{\dagger}(p_{1}) +
S(p_{1}, p_{2})\,  \B_{a}^{\dagger}(p_{2})\, \A_{b}^{\dagger}(p_{1}) 
\,, \label{ZFAB}
\ee
where
\be
T(p_{1}, p_{2}) &=&  \frac{1}{2} \tilde S_{0} (A-B) \rightarrow 
\frac{u_{1}-u_{2}}{u_{1}-u_{2}+i} \,, \non \\ 
S(p_{1}, p_{2}) &=&  \frac{1}{2} \tilde S_{0} (A+B) \rightarrow 
\frac{i}{u_{1}-u_{2}+i} \,,
\ee
which agrees with (\ref{SABphibarphi}).\footnote{There is a sign
discrepancy in $S(p_{1}, p_{2})$.  However, the sign of $S(p_{1},
p_{2})$ in (\ref{ZFAB}) can be changed by a gauge transformation, e.g.
by changing $\A_{1} \rightarrow -\A_{1}$ and leaving $\A_{2}\,,
\B_{1}\,, \B_{2}$ unchanged.}

\section{Discussion}\label{sec:discussion}

We have found that the all-loop $AdS_{4}/CFT_{3}$ $S$-matrix
(\ref{SAA}) - (\ref{SBA}) correctly reproduces the ${\cal N}=6$ CS
two-loop scalar-sector scattering amplitudes.  The scalar factors
(\ref{S0values}), which differ from the $AdS_{5}/CFT_{4}$ scalar
factor (\ref{AFrelation}), play a crucial role.  In particular, we
have confirmed that the scattering of $\A$ and $\B$ particles is
reflectionless.  This gives greater confidence in the correctness of
the all-loop $S$-matrix, and in the corresponding all-loop BAEs
\cite{GV2}.

We have restricted our analysis to the scalar sector of ${\cal N}=6$
CS, since this is the only sector for which an explicit Hamiltonian
has been available \cite{MZ2}.  Very recently, the Hamiltonian for the
full two-loop $OSp(6|4)$ spin chain has been found \cite{Zw, MSZ}.
Hence, it should now be possible to extend the present analysis to
other sectors, and thereby further check the all-loop $S$-matrix.

It would also be interesting to extend the present analysis beyond two
loops.  This could provide further information about the important
function $h(\lambda)$ (\ref{gvalue}) and the dressing phase in the
$S$-matrix.  However, such an analysis must wait until the higher-loop
Hamiltonian becomes available.

\section*{Acknowledgments}
C.A. thanks the University of Miami for hospitality during the course
of this work.  This work was supported in part by KRF-2007-313-C00150
and WCU grant R32-2008-000-10130-0 (C.A.), and by the National Science
Foundation under Grants PHY-0244261 and PHY-0554821 (R.N.).

\begin{appendix}

\section{Details of $\A-\B$ scattering}\label{sec:details}

In order to determine the $\A-\B$ scattering amplitudes, it is 
necessary to act with the Hamiltonian $H$ (\ref{HCS}) on the state 
(\ref{ABeigenket}). We catalog here the action of $H$ on the various
terms:
\be
H |x_{1}\,, x_{2} \rangle^{\A\B}_{\phi_{i}, \phi_{j}} &=& 
4|x_{1}\,, x_{2} \rangle^{\A\B}_{\phi_{i}, \phi_{j}} 
-|x_{1}-1\,, x_{2} \rangle^{\A\B}_{\phi_{i}, \phi_{j}} 
-|x_{1}+1\,, x_{2} \rangle^{\A\B}_{\phi_{i}, \phi_{j}} \non \\
&-& |x_{1}\,, x_{2}-1 \rangle^{\A\B}_{\phi_{i}, \phi_{j}}
-|x_{1}\,, x_{2}+1 \rangle^{\A\B}_{\phi_{i}, \phi_{j}}  \quad  
\mbox{ for } x_{1} 
< x_{2} - 1 \,, 
\ee
\be 
H |x_{1}\,, x_{1}+1 \rangle^{\A\B}_{A_{2}, A_{2}^{\dagger}} &=& 
4 |x_{1}\,, x_{1}+1 \rangle^{\A\B}_{A_{2}, A_{2}^{\dagger}}
-  |x_{1}\rangle_{A_{2} A_{2}^{\dagger}} 
-  |x_{1}+1\rangle_{A_{2} A_{2}^{\dagger}} \non \\
&-& |x_{1}-1\,, x_{1}+1 \rangle^{\A\B}_{A_{2}, A_{2}^{\dagger}}
- |x_{1}\,, x_{1}+2 \rangle^{\A\B}_{A_{2}, A_{2}^{\dagger}} \,, 
\label{a2a2d}
\ee
\be
H |x_{1}\,, x_{1}+1 \rangle^{\A\B}_{B_{2}^{\dagger}, B_{2}} &=& 
4 |x_{1}\,, x_{1}+1 \rangle^{\A\B}_{B_{2}^{\dagger}, B_{2}}
-  |x_{1}\rangle_{B_{2}^{\dagger} B_{2}} 
-  |x_{1}+1\rangle_{B_{2}^{\dagger} B_{2}} \non \\
&-& |x_{1}-1\,, x_{1}+1 \rangle^{\A\B}_{B_{2}^{\dagger}, B_{2}}
- |x_{1}\,, x_{1}+2 \rangle^{\A\B}_{B_{2}^{\dagger}, B_{2}} \,, 
\label{b2db2}
\ee
\be 
H |x_{1}\,, x_{1}+1 \rangle^{\A\B}_{A_{2}^{\dagger}, A_{2}} &=& 
4 |x_{1}\,, x_{1}+1 \rangle^{\A\B}_{A_{2}^{\dagger}, A_{2}}
- \frac{1}{2}|x_{1} \rangle_{A_{2}^{\dagger} A_{2}}
- \frac{1}{2}|x_{1}+1 \rangle_{A_{2} A_{2}^{\dagger}} \non \\
&+& \frac{1}{2}|x_{1} \rangle_{A_{1} A_{1}^{\dagger}}
+ \frac{1}{2}|x_{1}+1 \rangle_{A_{1} A_{1}^{\dagger}} 
+ \frac{1}{2}|x_{1} \rangle_{B_{1}^{\dagger} B_{1}}
+ \frac{1}{2}|x_{1}+1 \rangle_{B_{1}^{\dagger} B_{1}} \non \\
&+&  \frac{1}{2}|x_{1} \rangle_{B_{2}^{\dagger} B_{2}}
+ \frac{1}{2}|x_{1}+1 \rangle_{B_{2}^{\dagger} B_{2}} \non \\ 
&-& |x_{1}-1\,, x_{1}+1 \rangle^{\A\B}_{A_{2}^{\dagger}, A_{2}}
- |x_{1}\,, x_{1}+2 \rangle^{\A\B}_{A_{2}^{\dagger}, A_{2}} \,, 
\label{a2da2}
\ee
\be
H |x_{1}\,, x_{1}+1 \rangle^{\A\B}_{B_{2}, B_{2}^{\dagger}} &=& 
4 |x_{1}\,, x_{1}+1 \rangle^{\A\B}_{B_{2}, B_{2}^{\dagger}} 
- \frac{1}{2}|x_{1} \rangle_{B_{2}^{\dagger} B_{2}}
- \frac{1}{2}|x_{1}+1 \rangle_{B_{2}^{\dagger} B_{2}} \non \\
&+& \frac{1}{2}|x_{1} \rangle_{B_{1}^{\dagger} B_{1}}
+ \frac{1}{2}|x_{1}+1 \rangle_{B_{1}^{\dagger} B_{1}}
+ \frac{1}{2}|x_{1} \rangle_{A_{1} A_{1}^{\dagger}} 
+ \frac{1}{2}|x_{1}+1 \rangle_{A_{1} A_{1}^{\dagger}} \non \\
&+& \frac{1}{2}|x_{1} \rangle_{A_{2} A_{2}^{\dagger}}
+ \frac{1}{2}|x_{1}+1 \rangle_{A_{2} A_{2}^{\dagger}} \non \\
&-& |x_{1}-1\,, x_{1}+1 \rangle^{\A\B}_{B_{2}, B_{2}^{\dagger}}
- |x_{1}\,, x_{1}+2 \rangle^{\A\B}_{B_{2}, B_{2}^{\dagger}} \,, 
\label{b2b2d}
\ee
\be
H |x_{1} \rangle_{A_{1} A_{1}^{\dagger}} &=& 
3 |x_{1} \rangle_{A_{1} A_{1}^{\dagger}}
- \frac{1}{2} |x_{1}-1 \rangle_{A_{1} A_{1}^{\dagger}}
- \frac{1}{2} |x_{1}+1 \rangle_{A_{1} A_{1}^{\dagger}} \non \\
&+& \frac{1}{2} |x_{1} \rangle_{A_{2} A_{2}^{\dagger}}
+  \frac{1}{2} |x_{1}+1 \rangle_{A_{2} A_{2}^{\dagger}}
+ |x_{1} \rangle_{B_{1}^{\dagger} B_{1}}
+ |x_{1}+1 \rangle_{B_{1}^{\dagger} B_{1}}\non \\
&+& \frac{1}{2}|x_{1} \rangle_{B_{2}^{\dagger} B_{2}} 
+ \frac{1}{2}|x_{1}+1 \rangle_{B_{2}^{\dagger} B_{2}} 
+ \frac{1}{2}|x_{1}-1\,, x_{1}\rangle^{\A\B}_{A_{2}^{\dagger}, A_{2}} \non \\
&+& \frac{1}{2}|x_{1}\,, x_{1}+1\rangle^{\A\B}_{A_{2}^{\dagger}, A_{2}}
+\frac{1}{2}|x_{1}-1\,, x_{1} \rangle^{\A\B}_{B_{2}, B_{2}^{\dagger}}
+\frac{1}{2}|x_{1}\,, x_{1}+1 \rangle^{\A\B}_{B_{2}, B_{2}^{\dagger}} \,, 
\ee 
\be
H |x_{1} \rangle_{A_{2} A_{2}^{\dagger}} &=& 
4 |x_{1} \rangle_{A_{2} A_{2}^{\dagger}} 
+ \frac{1}{2} |x_{1}-1 \rangle_{A_{1} A_{1}^{\dagger}}
+ \frac{1}{2} |x_{1} \rangle_{A_{1} A_{1}^{\dagger}} \non \\
&+& \frac{1}{2}|x_{1} \rangle_{B_{1}^{\dagger} B_{1}}
+ \frac{1}{2}|x_{1}+1 \rangle_{B_{1}^{\dagger} B_{1}}
-|x_{1}-1\,, x_{1} \rangle^{\A\B}_{A_{2}, A_{2}^{\dagger}}
-|x_{1}\,, x_{1}+1 \rangle^{\A\B}_{A_{2}, A_{2}^{\dagger}} \non \\
&-& \frac{1}{2}|x_{1}-1\,, x_{1} \rangle^{\A\B}_{A_{2}^{\dagger}, A_{2}}
- \frac{1}{2}|x_{1}\,, x_{1}+1 \rangle^{\A\B}_{A_{2}^{\dagger}, 
A_{2}} \non \\
&+& \frac{1}{2}|x_{1}-1\,, x_{1} \rangle^{\A\B}_{B_{2}, B_{2}^{\dagger}}
+ \frac{1}{2}|x_{1}\,, x_{1}+1 \rangle^{\A\B}_{B_{2}, 
B_{2}^{\dagger}} \,, 
\ee
\be
H |x_{1} \rangle_{B_{1}^{\dagger} B_{1}} &=& 
3 |x_{1} \rangle_{B_{1}^{\dagger} B_{1}}
- \frac{1}{2} |x_{1}-1 \rangle_{B_{1}^{\dagger} B_{1}}
- \frac{1}{2} |x_{1}+1 \rangle_{B_{1}^{\dagger} B_{1}} \non \\
&+& \frac{1}{2} |x_{1}-1 \rangle_{B_{2}^{\dagger} B_{2}} 
+ \frac{1}{2} |x_{1} \rangle_{B_{2}^{\dagger} B_{2}} 
+ |x_{1}-1 \rangle_{A_{1} A_{1}^{\dagger}}
+ |x_{1} \rangle_{A_{1} A_{1}^{\dagger}} \non \\
&+& \frac{1}{2} |x_{1}-1 \rangle_{A_{2} A_{2}^{\dagger}} 
+ \frac{1}{2} |x_{1} \rangle_{A_{2} A_{2}^{\dagger}} 
+ \frac{1}{2}|x_{1}-1\,, x_{1} \rangle^{\A\B}_{A_{2}^{\dagger}, 
A_{2}} \non \\
&+& \frac{1}{2}|x_{1}\,, x_{1}+1 \rangle^{\A\B}_{A_{2}^{\dagger}, 
A_{2}}
+ \frac{1}{2}|x_{1}-1\,, x_{1} \rangle^{\A\B}_{B_{2}, B_{2}^{\dagger}}
+ \frac{1}{2}|x_{1}\,, x_{1}+1 \rangle^{\A\B}_{B_{2}, 
B_{2}^{\dagger}} \,, 
\ee
\be
H |x_{1} \rangle_{B_{2}^{\dagger} B_{2}} &=&
4 |x_{1} \rangle_{B_{2}^{\dagger} B_{2}} 
+ \frac{1}{2} |x_{1} \rangle_{B_{1}^{\dagger} B_{1}}
+ \frac{1}{2} |x_{1}+1 \rangle_{B_{1}^{\dagger} B_{1}}  \non \\
&+& \frac{1}{2} |x_{1}-1 \rangle_{A_{1} A_{1}^{\dagger}}
+ \frac{1}{2} |x_{1} \rangle_{A_{1} A_{1}^{\dagger}}
+ \frac{1}{2}|x_{1}-1 \,, x_{1} \rangle^{\A\B}_{A_{2}^{\dagger}, 
A_{2}} \non \\
&+& \frac{1}{2}|x_{1}\,, x_{1}+1 \rangle^{\A\B}_{A_{2}^{\dagger}, 
A_{2}}
- |x_{1}-1\,, x_{1} \rangle^{\A\B}_{B_{2}^{\dagger}, B_{2}}
- |x_{1}\,, x_{1}+1 \rangle^{\A\B}_{B_{2}^{\dagger}, B_{2}} \non \\
&-& \frac{1}{2} |x_{1}-1\,, x_{1} \rangle^{\A\B}_{B_{2}, B_{2}^{\dagger}}
- \frac{1}{2}|x_{1}\,, x_{1}+1 \rangle^{\A\B}_{B_{2}, 
B_{2}^{\dagger}} \,.
\ee
The appearance of terms of the form $|x\rangle_{A_{k}
A_{k}^{\dagger}}$ and $|x\rangle_{B_{k}^{\dagger} B_{k}}$ ($k = 1, 2$)
on the RHS of (\ref{a2a2d})-(\ref{b2b2d}) explains the need for such
terms in the eigenstate (\ref{ABeigenket}).

With the help of the above results, the eigenvalue equation
\be
H |\psi \rangle = E |\psi \rangle
\ee
with $|\psi \rangle$ and $E$ given by (\ref{ABeigenket}) and (\ref{2particleenergy}), 
respectively, leads to the following equations for the amplitudes:
\be
0 &=& \left[3 - 
\frac{1}{2}(e^{i(p_{1}+p_{2})}+e^{-i(p_{1}+p_{2})}) - E \right] 
A_{A_{1}A_{1}^{\dagger}} \non \\
&+& (1 + e^{i(p_{1}+p_{2})})\left( \frac{1}{2} A_{A_{2}A_{2}^{\dagger}}
+ A_{B_{1}^{\dagger}B_{1}} + \frac{1}{2} A_{B_{2}^{\dagger}B_{2}} 
\right)  \non \\
&+&  \frac{1}{2}(e^{i p_{2}}+ e^{-i p_{1}}) \left[ A_{A_{2}^{\dagger}A_{2}}(12)
+ A_{B_{2} B_{2}^{\dagger}}(12) \right] \non \\
&+& \frac{1}{2}(e^{i p_{1}}+ e^{-i p_{2}}) \left[ 
A_{A_{2}^{\dagger}A_{2}}(21)
+ A_{B_{2} B_{2}^{\dagger}}(21) \right] \,,
\ee
\be
0 &=& \frac{1}{2}(1 + e^{-i(p_{1}+p_{2})}) A_{A_{1}A_{1}^{\dagger}}
+ (4 - E) A_{A_{2}A_{2}^{\dagger}} 
+ \frac{1}{2}(1 + e^{i(p_{1}+p_{2})}) A_{B_{1}^{\dagger}B_{1}} \non \\
&+& (e^{i p_{2}}+ e^{-i p_{1}}) \left[ 
- A_{A_{2} A_{2}^{\dagger}}(12)
- \frac{1}{2} A_{A_{2}^{\dagger}A_{2}}(12)
+ \frac{1}{2} A_{B_{2} B_{2}^{\dagger}}(12) \right] \non \\
&+& (e^{i p_{1}}+ e^{-i p_{2}}) \left[ 
- A_{A_{2} A_{2}^{\dagger}}(21)
- \frac{1}{2} A_{A_{2}^{\dagger}A_{2}}(21)
+ \frac{1}{2} A_{B_{2} B_{2}^{\dagger}}(21) \right] \,,
\ee
\be
0 &=& (1 + e^{-i(p_{1}+p_{2})}) \left[
A_{A_{1}A_{1}^{\dagger}}
+ \frac{1}{2} A_{A_{2}A_{2}^{\dagger}} 
+ \frac{1}{2} A_{B_{2}^{\dagger}B_{2}} \right] \non \\
&+& \left[3 - 
\frac{1}{2}(e^{i(p_{1}+p_{2})}+e^{-i(p_{1}+p_{2})}) - E \right] 
 A_{B_{1}^{\dagger}B_{1}}  \non \\
&+&  \frac{1}{2}(e^{i p_{2}}+ e^{-i p_{1}}) \left[ A_{A_{2}^{\dagger}A_{2}}(12)
+ A_{B_{2} B_{2}^{\dagger}}(12) \right] \non \\ 
&+& \frac{1}{2}(e^{i p_{1}}+ e^{-i p_{2}}) \left[ 
A_{A_{2}^{\dagger}A_{2}}(21)
+ A_{B_{2} B_{2}^{\dagger}}(21) \right] \,,
\ee
\be
0 &=& \frac{1}{2}(1 + e^{-i(p_{1}+p_{2})}) A_{A_{1}A_{1}^{\dagger}}
+ \frac{1}{2}(1 + e^{i(p_{1}+p_{2})}) A_{B_{1}^{\dagger}B_{1}} 
+ (4 - E)  A_{B_{2}^{\dagger} B_{2}} \non \\
&+& (e^{i p_{2}}+ e^{-i p_{1}}) \left[ 
- A_{B_{2}^{\dagger} B_{2}}(12)
+ \frac{1}{2} A_{A_{2}^{\dagger}A_{2}}(12)
- \frac{1}{2} A_{B_{2} B_{2}^{\dagger}}(12) \right] \non \\
&+& (e^{i p_{1}}+ e^{-i p_{2}}) \left[ 
- A_{B_{2}^{\dagger} B_{2}}(21)
+ \frac{1}{2} A_{A_{2}^{\dagger}A_{2}}(21)
- \frac{1}{2} A_{B_{2} B_{2}^{\dagger}}(21) \right] \,,
\ee
\be
0 &=& e^{i p_{2}} (4 - e^{-i p_{1}} - e^{i p_{2}} - E) 
A_{A_{2} A_{2}^{\dagger}}(12) 
+ e^{i p_{1}} (4 - e^{-i p_{2}} - e^{i p_{1}} - E) 
A_{A_{2} A_{2}^{\dagger}}(21) \non \\
&-& (1 + e^{i(p_{1}+p_{2})}) A_{A_{2} A_{2}^{\dagger}} \,,
\ee
\be
0 &=& e^{i p_{2}} (4 - e^{-i p_{1}} - e^{i p_{2}} - E) 
A_{B_{2}^{\dagger} B_{2}}(12) 
+ e^{i p_{1}} (4 - e^{-i p_{2}} - e^{i p_{1}} - E) 
A_{B_{2}^{\dagger} B_{2}}(21) \non \\
&-& (1 + e^{i(p_{1}+p_{2})}) A_{B_{2}^{\dagger} B_{2}} \,,
\ee 
\be
0 &=& e^{i p_{2}} (4 - e^{-i p_{1}} - e^{i p_{2}} - E) 
A_{A_{2}^{\dagger} A_{2}}(12) 
+ e^{i p_{1}} (4 - e^{-i p_{2}} - e^{i p_{1}} - E) 
A_{A_{2}^{\dagger} A_{2}}(21) \non \\
&+& \frac{1}{2}(1 + e^{i(p_{1}+p_{2})}) \left(
A_{A_{1}A_{1}^{\dagger}}-A_{A_{2} A_{2}^{\dagger}}
+ A_{B_{1}^{\dagger}B_{1}} + A_{B_{2}^{\dagger} B_{2}} \right) \,,
\ee
\be
0 &=& e^{i p_{2}} (4 - e^{-i p_{1}} - e^{i p_{2}} - E) 
A_{B_{2} B_{2}^{\dagger}}(12) 
+ e^{i p_{1}} (4 - e^{-i p_{2}} - e^{i p_{1}} - E) 
A_{B_{2} B_{2}^{\dagger}}(21) \non \\
&+& \frac{1}{2}(1 + e^{i(p_{1}+p_{2})}) \left(
A_{A_{1}A_{1}^{\dagger}}+A_{A_{2} A_{2}^{\dagger}}
+ A_{B_{1}^{\dagger}B_{1}} - A_{B_{2}^{\dagger} B_{2}} \right) \,.
\ee
Eliminating $A_{A_{k}A_{k}^{\dagger}}$, $A_{B_{k}^{\dagger}B_{k}}$
($k=1, 2$), and then solving for the $(21)$ amplitudes in terms of the
$(12)$ amplitudes, we arrive at the results (\ref{inout}),
(\ref{SABphibarphi}).

\end{appendix}


\begin{thebibliography}{99}

\bibitem{ZZ}
A.~B. Zamolodchikov and Al.~B. Zamolodchikov, 
``Factorized $S$ matrices in two-dimensions as the exact solutions 
of certain relativistic quantum field models,'' 
{\it Ann.  Phys.}  {\bf 120}, 253 (1979).  

\bibitem{AdSCFT}
J.~M. Maldacena, ``The large $N$ limit of superconformal
field theories and supergravity,'' 
{\it Adv. Theor. Math. Phys.} {\bf 2}, 231 (1998) 
[arXiv:hep-th/9711200].\
$\bullet$ S.~S. Gubser, I.~R. Klebanov and A.~M. Polyakov, 
``Gauge theory correlators from non-critical string theory,''
{\it Phys. Lett.} {\bf B428}, 105 (1998) 
[arXiv:hep-th/9802109].\
$\bullet$ E. Witten, 
``Anti-de Sitter space and holography,'' 
{\it Adv. Theor. Math. Phys.} {\bf 2}, 253 (1998) 
[arXiv:hep-th/9802150].
    
\bibitem{MZ1}
J.~A. Minahan and K. Zarembo,
``The Bethe-Ansatz for ${\cal N}=4$ Super Yang-Mills,''
{\it JHEP} {\bf 0303}, 013 (2003)   
[arXiv:hep-th/0212208].

\bibitem{BS1}
N. Beisert and M. Staudacher,
``The ${\cal N}=4$ SYM Integrable Super Spin Chain,''
{\it Nucl. Phys.} {\bf B670}, 439 (2003) 
[arXiv:hep-th/0307042].

\bibitem{BV}
D. Berenstein and S.~E. V\'azquez,
``Integrable open spin chains from giant gravitons,''
{\it JHEP} {\bf 0506}, 059 (2005)
[arXiv:hep-th/0501078].

\bibitem{reviews}
A.~A. Tseytlin,
``Spinning strings and AdS/CFT duality,''
in Ian Kogan Memorial Volume, {\it From Fields to Strings: Circumnavigating
Theoretical Physics}, M. Shifman, A. Vainshtein, and J. Wheater, eds.
(World Scientific, 2004)
[arXiv:hep-th/0311139].\
$\bullet$ N. Beisert,
``The dilatation operator of ${\cal N} = 4$ super Yang-Mills theory and
integrability,''
{\it Phys. Rept.}  {\bf 405}, 1 (2005)
[arXiv:hep-th/0407277].\
$\bullet$ K. Zarembo,
``Semiclassical Bethe ansatz and AdS/CFT,''
{\it Comptes Rendus Physique} {\bf 5}, 1081 (2004)
[{\it Fortsch. Phys.}  {\bf 53}, 647 (2005)]
[arXiv:hep-th/0411191].\
$\bullet$ J. Plefka,
``Spinning strings and integrable spin chains in the AdS/CFT
correspondence,''
{\it Living Rev. Rel.}  {\bf 8}, 9 (2005)
[arXiv:hep-th/0507136].\
$\bullet$ J.~A. Minahan, 
``A brief introduction to the Bethe ansatz in ${\cal N}=4$ 
super-Yang-Mills,''
{\it J. Phys.} {\bf A39}, 12657 (2006).\
$\bullet$ K. Okamura,
``Aspects of Integrability in AdS/CFT Duality,''
[arXiv:0803.3999].

\bibitem{St}
M. Staudacher,
``The factorized $S$-matrix of CFT/AdS,''
{\it JHEP} {\bf 0505}, 054 (2005)
[arXiv:hep-th/0412188].
    
\bibitem{Be}
N. Beisert,
``The $su(2|2)$ dynamic $S$-matrix,''
{\it Adv. Theor. Math. Phys.}  {\bf 12}, 945 (2008)
[arXiv:hep-th/0511082].\
$\bullet$ N. Beisert,
``The Analytic Bethe Ansatz for a Chain with Centrally Extended
$su(2|2)$ Symmetry,''
{\it J. Stat. Mech.}  {\bf 0701}, P017 (2007)
[arXiv:nlin/0610017].

\bibitem{Ja}
R.~A. Janik,
``The $AdS_{5} \times S^{5}$ superstring worldsheet $S$-matrix and crossing 
symmetry,''
{\it Phys. Rev.} {\bf D73}, 086006 (2006)
[arXiv:hep-th/0603038].

\bibitem{AF1}
G. Arutyunov and S. Frolov,
``On $AdS_{5} \times S^{5}$ string $S$-matrix,''
{\it Phys. Lett.  } {\bf B639}, 378 (2006)
[arXiv:hep-th/0604043].

\bibitem{BHL}
N. Beisert, R. Hernandez and E. Lopez,
``A crossing-symmetric phase for $AdS_{5} \times S^{5}$ strings,''
{\it JHEP} {\bf 0611}, 070 (2006)
[arXiv:hep-th/0609044].

\bibitem{BES}
N. Beisert, B. Eden and M. Staudacher,
``Transcendentality and crossing,''
{\it J. Stat. Mech.}  {\bf 0701}, P021 (2007)
[arXiv:hep-th/0610251].

\bibitem{AFZ}
G. Arutyunov, S. Frolov and M. Zamaklar,
`The Zamolodchikov-Faddeev algebra for $AdS_{5} \times S^{5}$ superstring,''
{\it JHEP} {\bf 0704}, 002 (2007)
[arXiv:hep-th/0612229].

\bibitem{DHM}
N. Dorey, D.M. Hofman and J.M. Maldacena,
``On the singularities of the magnon S-matrix,''
{\it Phys. Rev.  } {\bf D76}, 025011 (2007)
[arXiv:hep-th/0703104].

\bibitem{MM}
M.~J. Martins and C.S. Melo,
``The Bethe ansatz approach for factorizable centrally extended
$S$-matrices,''
{\it Nucl. Phys.} {\bf B785}, 246 (2007)
[arXiv:hep-th/0703086].

\bibitem{dL}
M. de Leeuw,
``Coordinate Bethe Ansatz for the String $S$-Matrix,''
{\it J. Phys. } {\bf A40}, 14413 (2007)
[arXiv:0705.2369].

\bibitem{BS2}
N. Beisert and M. Staudacher,
``Long-range $PSU(2,2|4)$ Bethe ansaetze for gauge theory and strings,''
{\it Nucl. Phys.  } {\bf B727}, 1 (2005)
[arXiv:hep-th/0504190].

\bibitem{ABJM}
O. Aharony, O. Bergman, D.~L. Jafferis and J. Maldacena,
``${\cal N}=6$ superconformal Chern-Simons-matter theories, M2-branes and their
gravity duals,''
{\it JHEP} {\bf 0810}, 091 (2008)
[arXiv:0806.1218].

\bibitem{MZ2}
J.~A. Minahan and K. Zarembo,
``The Bethe ansatz for superconformal Chern-Simons,''
{\it JHEP} {\bf 0809}, 040 (2008)
[arXiv:0806.3951]

\bibitem{BR}
D. Bak and S.~J. Rey,
``Integrable Spin Chain in Superconformal Chern-Simons Theory,''
{\it JHEP} {\bf 0810}, 053 (2008)
[arXiv:0807.2063]

\bibitem{AF2} 
G. Arutyunov and S. Frolov,
``Superstrings on $AdS_4 \times CP^3$ as a Coset Sigma-model,'' 
{\it JHEP} {\bf 0809}, 129 (2008)
[arXiv:0806.4940].

\bibitem{Stef} 
B.~J. Stefanski,
``Green-Schwarz action for Type IIA strings on $AdS_4\times CP^3$,''
[arXiv:0806.4948].

\bibitem{GV1}
N. Gromov and P. Vieira,
``The AdS4/CFT3 algebraic curve,'' 
[arXiv:0807.0437].

\bibitem{GV2} 
N. Gromov and P. Vieira,
``The all loop AdS4/CFT3 Bethe ansatz,'' 
{\it JHEP} {\bf 0901}, 016 (2009)
[arXiv:0807.0777].

\bibitem{NT}
T. Nishioka and T. Takayanagi,
``On Type IIA Penrose Limit and ${\cal N}=6$ Chern-Simons Theories,''
{\it JHEP} {\bf 0808}, 001 (2008)
[arXiv:0806.3391].

\bibitem{GGY}
D. Gaiotto, S. Giombi and X. Yin,
``Spin Chains in ${\cal N}=6$ Superconformal Chern-Simons-Matter Theory,''
[arXiv:0806.4589].

\bibitem{GHO}
G. Grignani, T. Harmark and M. Orselli,
``The $SU(2) \times SU(2)$ sector in the string dual of ${\cal N}=6$ superconformal
Chern-Simons theory,''
[arXiv:0806.4959].

\bibitem{AN1}
C. Ahn and R.~I. Nepomechie,
``${\cal N}=6$ super Chern-Simons theory $S$-matrix and all-loop Bethe ansatz
equations,''
{\it JHEP} {\bf 0809}, 010 (2008)
[arXiv:0807.1924].

\bibitem{AN2}
C. Ahn and R.~I. Nepomechie,
``An alternative $S$-matrix for ${\cal N}=6$ Chern-Simons theory ?''
[arXiv:0810.1915].

\bibitem{more}
 M. Benna, I. Klebanov, T. Klose and M. Smedback,
``Superconformal Chern-Simons Theories and $AdS_4/CFT_3$
Correspondence,'' {\it JHEP} {\bf 0809}, 072 (2008)
[arXiv:0806.1519].\
$\bullet$ K.~Hanaki and H.~Lin,
``M2-M5 Systems in ${\cal N}=6$ Chern-Simons Theory,''
{\it JHEP} {\bf 0809}, 067 (2008)
[arXiv:0807.2074].\
$\bullet$ I.~Shenderovich, 
``Giant magnons in $AdS_4/CFT_3$:
dispersion, quantization and finite--size corrections,''
[arXiv:0807.2861].\ 
$\bullet$ C.~Ahn, P.~Bozhilov and R.~C.~Rashkov,
``Neumann-Rosochatius integrable system for strings on $AdS_4 \times
CP^3$,'' {\it JHEP} {\bf 0809}, 017 (2008) [arXiv:0807.3134].\
$\bullet$ Y.~Honma, S.~Iso, Y.~Sumitomo,
H.~Umetsu and S.~Zhang, ``Generalized Conformal Symmetry and
Recovery of $SO(8)$ in Multiple M2 and D2 Branes,''
[arXiv:0807.3825].\ 
$\bullet$ T.~McLoughlin and R.~Roiban,
``Spinning strings at one-loop in $AdS_4 \times P^3$,'' {\it JHEP}
{\bf 0812}, 101 (2008) [arXiv:0807.3965].\ 
$\bullet$ J.~Kluson, 
``D2 to M2 Procedure for D2-Brane DBI Effective Action,'' 
{\it Nucl.\ Phys.} {\bf B808}, 260 (2009) [arXiv:0807.4054].\ 
$\bullet$ L.~F.~Alday, G.~Arutyunov and D.~Bykov, 
``Semiclassical Quantization
of Spinning Strings in $AdS_4 \times CP^3$,'' {\it JHEP} {\bf 0811},
089 (2008) [arXiv:0807.4400].\ 
$\bullet$ C.~Krishnan, ``AdS4/CFT3 at
One Loop,'' {\it JHEP} {\bf 0809}, 092 (2008) [arXiv:0807.4561].\
$\bullet$ N.~Gromov and V.~Mikhaylov, 
``Comment on the Scaling
Function in $AdS4 \times CP3$,'' [arXiv:0807.4897].\ 
$\bullet$ O.~Aharony, O.~Bergman and D.~L.~Jafferis, 
``Fractional M2-branes,'' {\it JHEP}
{\bf 0811}, 043 (2008) [arXiv:0807.4924].\ 
$\bullet$ G.~Bonelli,
A.~Tanzini and M.~Zabzine, 
``Topological branes, p-algebras and
generalized Nahm equations,'' [arXiv:0807.5113].\ 
$\bullet$
M.~R.~Garousi and A.~Ghodsi, 
``Hydrodynamics of ${\cal N}=6$ Superconformal
Chern-Simons Theories at Strong Coupling,'' [arXiv:0808.0411].\
$\bullet$  R.~D'Auria, P.~Fre, P.~A.~Grassi and M.~Trigiante,
``Superstrings on $AdS_4 \times CP^3$ from Supergravity,''
[arXiv:0808.1282].\ 
$\bullet$  D.~Fioravanti, P.~Grinza and M.~Rossi, 
``The generalised scaling function: a systematic study,''
[arXiv:0808.1886].\ 
$\bullet$  D.~Berenstein and D.~Trancanelli,
``Three-dimensional ${\cal N}=6$ SCFT's and their membrane dynamics,''
[arXiv:0808.2503].\ 
$\bullet$ R.~C.~Rashkov, 
``A note on the
reduction of the $AdS4 \times CP3$ string sigma model,'' {\it Phys.\ Rev.}
{\bf D78}, 106012 (2008) [arXiv:0808.3057].\ 
$\bullet$ K.~Hosomichi,
K.~M.~Lee, S.~Lee, S.~Lee, J.~Park and P.~Yi, 
``A Nonperturbative
Test of M2-Brane Theory,'' {\it JHEP} {\bf 0811}, 058 (2008)
[arXiv:0809.1771].\ 
$\bullet$ J.~Kluson and K.~L.~Panigrahi,
``Defects and Wilson Loops in 3d QFT from D-branes in $AdS(4) \times
CP^{3}$,'' [arXiv:0809.3355].\ 
$\bullet$ C.-h.~Ahn, 
``Squashing
Gravity Dual of ${\cal N}=6$ Superconformal Chern-Simons Gauge Theory,''
[arXiv:0809.3684].\ 
$\bullet$  T.~McLoughlin, R.~Roiban and A.~A.~Tseytlin, 
``Quantum spinning strings in $AdS_4 \times CP^3$:
testing the Bethe Ansatz proposal,'' {\it JHEP} {\bf 0811}, 069
(2008) [arXiv:0809.4038].\ 
$\bullet$ S.~Ryang, 
``Giant Magnon and
Spike Solutions with Two Spins in $AdS4 \times CP3$,'' {\it JHEP} {\bf
0811}, 084 (2008) [arXiv:0809.5106].\ 
$\bullet$  D.~Bombardelli and D.~Fioravanti, 
``Finite-Size Corrections of the $\mathbb{CP}^3$
Giant Magnons: the L\'{u}scher terms,'' [arXiv:0810.0704].\
$\bullet$ T.~Lukowski and O.~O.~Sax, 
``Finite size giant magnons in
the $SU(2) \times SU(2)$ sector of $AdS_4 \times CP^3$,'' {\it JHEP} {\bf
0812}, 073 (2008) [arXiv:0810.1246].\ 
$\bullet$  M.~Kreuzer, R.~C.~Rashkov and M.~Schimpf, 
``Near Flat Space limit of strings on
$AdS_4 \times CP^3$,'' [arXiv:0810.2008].\ 
$\bullet$  C.~Ahn and P.~Bozhilov, 
``Finite-size Effect of the Dyonic Giant Magnons in ${\cal N}=6$
super Chern-Simons Theory,'' [arXiv:0810.2079].\ 
$\bullet$ C.~Ahn and P.~Bozhilov, 
``M2-brane Perspective on ${\cal N}=6$ Super Chern-Simons
Theory at Level k,'' {\it JHEP} {\bf 0812}, 049 (2008)
[arXiv:0810.2171].\ 
$\bullet$ F.~Spill, 
``Weakly coupled ${\cal N}=4$ Super
Yang-Mills and ${\cal N}=6$ Chern-Simons theories from $u(2|2)$ Yangian
symmetry,'' [arXiv:0810.3897].\ 
$\bullet$  J.~Gomis, D.~Sorokin and L.~Wulff, 
``The complete $AdS(4) \times CP(3)$ superspace for the type IIA
superstring and D-branes,'' [arXiv:0811.1566].\ 
$\bullet$
C.~Kristjansen, M.~Orselli and K.~Zoubos, 
``Non-planar ABJM Theory
and Integrability,'' [arXiv:0811.2150].\ 
$\bullet$  P.~Sundin, 
``The
$AdS(4) \times CP(3)$ string and its Bethe equations in the near plane wave
limit,'' [arXiv:0811.2775].\ 
$\bullet$ J.~H.~Baek, S.~Hyun, W.~Jang and S.~H.~Yi, 
``Membrane Dynamics in Three dimensional ${\cal N}=6$
Supersymmetric Chern-Simons Theory,'' [arXiv:0812.1772].\ 
$\bullet$ D.~Bak, 
``Zero Modes for the Boundary Giant Magnons,''
[arXiv:0812.2645].\ 
$\bullet$ A.~Agarwal, N.~Beisert and T.~McLoughlin, 
``Scattering in Mass-Deformed ${\cal N}\ge 4$ Chern-Simons
Models,'' [arXiv:0812.3367].

\bibitem{Zw}
B.~I. Zwiebel,
``Two-loop Integrability of Planar ${\cal N}=6$ Superconformal Chern-Simons Theory,''
[arXiv:0901.0411].

\bibitem{MSZ}
J.~A. Minahan, W. Schulgin and K. Zarembo,
``Two loop integrability for Chern-Simons theories with ${\cal N}=6$ supersymmetry,''
[arXiv:0901.1142].

\end{thebibliography}
\end{document}